\documentstyle[12pt]{article}
\title{Fundamental CP-violating quantities
in a SU(2)$ \otimes $U(1) model with many Higgs doublets}
\author{L.\ Lavoura\thanks{On leave of absence
from Universidade T\'ecnica de Lisboa,
Lisbon, Portugal}\hspace{1.5mm} and Jo\~ao P.\ Silva \\
\small Department of Physics, Carnegie-Mellon University, \\
\small Pittsburgh, Pennsylvania 15213, U.S.A.}
\begin{document}
\maketitle
\begin{abstract}
We consider a SU(2)$ \otimes $U(1) gauge theory
with many scalar doublets,
but without fermions.
We devise a systematic method of constructing quantities
which
are invariant under changes of basis of the scalar fields,
but acquire a minus sign under a general CP transformation.
Those quantities provide a basis-independent way of checking
for the existence of CP violation.
\end{abstract}

\vspace{5mm}

\section{Introduction}

In 1985,
Jarlskog \cite{jarlskog} and,
almost simultaneously,
Dunietz,
Greenberg,
and Wu \cite{wu}
discovered that CP violation in the quark sector
of the three-generation standard model originates in a single basic
CP-violating quantity,
\begin{eqnarray}
J & \equiv & \det [M_u M_u^{\dagger}, M_d M_d^{\dagger}]
\nonumber\\
  & \propto &
(m_t^2 - m_c^2)
(m_t^2 - m_u^2)
(m_c^2 - m_u^2)
(m_b^2 - m_s^2)
(m_b^2 - m_d^2)
(m_s^2 - m_d^2)
\nonumber\\
  &  &
\times {\rm Im} (V_{ud} V_{cs} V_{us}^{\ast} V_{cd}^{\ast})\, ,
\label{eq:J}
\end{eqnarray}
where $ M_u $ and $ M_d $ are the mass matrices
of the up-type and down-type quarks,
respectively,
and $ V $ is the Cabibbo--Kobayashi--Maskawa matrix.
Shortly afterwards,
Bernab\'eu,
Branco,
and Gronau (BBG) \cite{branco}
found $ J $ by searching for quantities which,
though invariant under a basis transformation of the fermion fields,
acquire a minus sign under a CP transformation.
Their method of deriving basic CP-odd quantities
in the fermion--gauge sector of the standard model
works for any number of fermion generations,
though unfortunately for more than three generations
a minimal set of independent CP-odd quantities is difficult to find
\cite{gronau}.

The methods of BBG have later been used in the construction
of basic CP-odd quantities
in models with vector-like quarks \cite{lavoura},
with left-right symmetry \cite{rebelo},
with neutrinos with Majorana masses \cite{nesbitt},
and in supersymmetric models \cite{kostelecky}.
With different aims,
general CP transformations have also been extensively studied
by Grimus and collaborators \cite{grimus}.

Recently,
M\'endez and Pomarol \cite{pomarol}
studied CP violation in the scalar sector
of a SU(2)$ \otimes $U(1) model with many scalar multiplets.
They only considered the possible clash
between the CP transformation properties
of the scalar gauge interactions and of the scalar mass matrices.
In particular,
for the simple case of the two-Higgs-doublet model,
they phenomenologically derived the existence
of a basic CP-violating quantity $ J_1 $,
analogous to the $ J $ of Eq.~\ref{eq:J}.

In this work we consider the derivation of $ J_1 $
as a basic CP-odd quantity under a general CP transformation,
along the same line of reasoning by which
BBG found $ J $ from basic principles.
We further consider the CP violation
originating not only in the mass matrices of the scalars,
but also in the cubic and quartic scalar interactions,
and do so for any number of scalar doublets.
For simplicity,
we only study a SU(2)$ \otimes $U(1) model with many scalar doublets,
and do not consider the possibilities of either an extended gauge group,
or of scalars in other representations of the gauge group.
We also do not consider the Yukawa interactions
of the scalars with the fermions,
though in general these will constitute further sources of CP violation.
This we do in order to keep our study within the limits
of reasonable generality.
However,
our methods,
supplemented by the ones of BBG,
can,
in principle,
be applied to CP violation in specific models
with both scalars and fermions.

In section 2 we explain our method
for the neutral sector of the two-Higgs-doublet model.
In section 3 we extend the method to the multi-Higgs-doublet case.
In section 4 we extend it to the charged sector.
Section 5 summarizes our results.

\section{Two-doublet case}

We consider a SU(2)$ \otimes $U(1) gauge model
with two scalar doublets $ H_1 $ and $ H_2 $.
We do not impose any symmetry beyond the gauge symmetry.
The most general Higgs potential consistent with renormalizability is
\begin{eqnarray}
V & = &
\mu_1 H_1^{\dagger} H_1 + \mu_2 H_2^{\dagger} H_2
+ (\mu_3 H_1^{\dagger} H_2 + h.c.)
\nonumber\\
  &   &
+ \lambda_1 (H_1^{\dagger} H_1)^2
+ \lambda_2 (H_2^{\dagger} H_2)^2
+ \lambda_3 (H_1^{\dagger} H_1) (H_2^{\dagger} H_2)
+ \lambda_4 (H_1^{\dagger} H_2) (H_2^{\dagger} H_1)
\nonumber\\
  &   &
+ \left[ \lambda_5 (H_1^{\dagger} H_2)^2
+ \lambda_6 (H_1^{\dagger} H_1) (H_1^{\dagger} H_2)
+ \lambda_7 (H_2^{\dagger} H_2) (H_1^{\dagger} H_2)
+ h.c.\right]\, ,
\label{eq:potential}
\end{eqnarray}
in which all the coupling constants,
except $ \mu_3 $,
$ \lambda_5 $,
$ \lambda_6 $,
and $ \lambda_7 $,
are real by hermiticity.
Without loss of generality,
we assume that the two scalar doublets are in the basis
in which only one of them,
$ H_1 $,
has a non-zero vacuum expectation value (VEV),
$ v $,
which is real.
Thus,
\begin{eqnarray}
H_1 & = &
\left( \begin{array}{c}
G^+ \\ v + (H^0 + i G^0)/\sqrt{2}
\end{array} \right)\, ,
\label{eq:H1}\\
H_2 & = &
\left( \begin{array}{c}
H^+ \\ (R + i I)/\sqrt{2}
\end{array} \right)\, ,
\label{eq:H2}
\end{eqnarray}
where $ H^0 $,
$ R $ and $ I $ are real neutral fields,
and $ G^+ $ and $ G^0 $ are the Goldstone bosons,
which,
in the unitary gauge,
become the longitudinal components of the $ W^+ $ and of the $ Z^0 $.
The stability conditions of the vacuum read
\begin{eqnarray}
\mu_1 = - 2 \lambda_1 v^2\, ,
\label{eq:mu1}\\
\mu_3 = - \lambda_6 v^2\, .
\label{eq:mu3}
\end{eqnarray}
We use these conditions to eliminate $ \mu_1 $ and $ \mu_3 $
as independent variables from $ V $.
The fact that Eqs.~\ref{eq:mu1} and \ref{eq:mu3}
appear to overdetermine $ v $
is connected to the fact that we have
chosen the basis in which $ H_2 $ has vanishing VEV.
This does not represent any loss of generality.

Let us consider the potential $ V $ as a function
of the fields $ \xi_1 \equiv R $ and $ \xi_2 \equiv I $.
$ V $ can be written as a sum in which some terms are of
one of the following forms\footnote{Other terms,
including the CP-odd field $G^0$, such as
$G^0 H^0 \xi_i$, will be discussed later.
Terms including the charged scalars, such as $G^- H^+ \xi_i$,
will be considered in section 4,
when we discuss the treatment of quantities
involving the charged fields.}:
\begin{enumerate}
\item $ a S $,
in which $ S $ denotes a general CP even monomial of
$ H^0 $,
$ (G^0 G^0)$,
$ (G^- G^+) $ and $ (H^- H^+) $
(or a constant),
and $ a $ denotes the respective coefficient,
which is a polynomial of the coupling constants in $ V $;
\item $ b_i \xi_i S $,
terms which are linear in either $ R $ or $ I $;
the $ b_i $ are two coupling constants\footnote{We use
throughout this paper the summation convention.};
\item $ c_{ij} \xi_i \xi_j S $,
terms quadratic in $ R $ and $ I $,
where the couplings $ c_{ij} $ can be viewed as a symmetric
$ 2 \times 2 $ matrix;
\item $ d_{ijk} \xi_i \xi_j \xi_k S $,
where the couplings $ d_{ijk} $
are symmetric under the interchange of any pair of indices;
\item $ e_{ijkl} \xi_i \xi_j \xi_k \xi_l S $,
with $ e_{ijkl} $ totally symmetric.
\end{enumerate}
To be exact, the coefficients $a$ through $e_{ijkl}$ should
carry a superscript $S$ specifying which monomial they refer to
[for example,
$ S=1 $,
or $S = H^0$,
or $ S=(H^0)^2$,
or $ S=H^0 H^- H^+$].
We will suppress that superscript
in order not to clutter the notation,
and will instead introduce primes in some specific examples.

There is in $ V $,
for instance,
a cubic interaction
\begin{equation}
\sqrt{2} v H^- H^+ \left[
\lambda_3 H^0 + ({\rm Re} \lambda_7) R - ({\rm Im} \lambda_7) I
\right]\, ,
\label{eq:chargedinteraction}
\end{equation}
which can be considered as the sum of a term of the form 1,
with $ S = H^0 H^- H^+ $ and $ a^{\prime} = \sqrt{2} v \lambda_3 $,
and of a term of the form 2,
with $ S = H^- H^+ $,
and
$ b^{\prime}_1 = \sqrt{2} v {\rm Re} \lambda_7 $,
$ b^{\prime}_2 = - \sqrt{2} v {\rm Im} \lambda_7 $.

The quadratic terms
\begin{equation}
\frac{1}{2} \left( \begin{array}{ccc} H^0 & R & I \end{array} \right)
M \left( \begin{array}{c} H^0 \\ R \\ I \end{array} \right)\, ,
\label{eq:massterms}
\end{equation}
where $ M $ is a $ 3 \times 3 $ real symmetric matrix,
are of particular interest,
because they are the mass terms of the neutral scalars.
By diagonalizing $ M $,
\begin{equation}
T M T^T = {\rm diag} (m_1^2, m_2^2, m_3^2)\, ,
\label{eq:diagonalization}
\end{equation}
where $ T $ is a matrix of SO(3),
one obtains the squared masses $ m_i^2 $
of the three physical scalars $ X_i $:
\begin{equation}
\left( \begin{array}{c} X_1 \\ X_2 \\ X_3 \end{array} \right)
=
T\, \left( \begin{array}{c} H \\ R \\ I \end{array} \right)\, .
\label{eq:physicalscalars}
\end{equation}
The mass terms in Eq.~\ref{eq:massterms} can be viewed as
the sum of a term of the form 1,
$ (1/2) M_{11} (H^0)^2 $,
a term of the form 2,
$ H^0 (M_{12} R + M_{13} I) $,
and a term of the form 3,
$ (1/2) (M_{22} R^2 + 2 M_{23} R I + M_{33} I^2) $.
The specific values of the coefficients are
\begin{eqnarray}
M_{11} & = & 4 \lambda_1 v^2\, ,
\label{eq:M11}\\
M_{22} & = & \mu_2 + (\lambda_3 + \lambda_4 + 2 {\rm Re} \lambda_5) v^2\, ,
\label{eq:M22}\\
M_{33} & = & \mu_2 + (\lambda_3 + \lambda_4 - 2 {\rm Re} \lambda_5) v^2\, ,
\label{eq:M33}\\
M_{23} & = & - 2 v^2 {\rm Im} \lambda_5\, ,
\label{eq:M23}\\
M_{12} & = & 2 v^2 {\rm Re} \lambda_6\, ,
\label{eq:M12}\\
M_{13} & = & - 2 v^2 {\rm Im} \lambda_6\, .
\label{eq:M13}
\end{eqnarray}

Now consider a basis transformation.
We do not want to leave the useful basis
in which only $ H_1 $ has a VEV,
and that VEV is real;
therefore,
the most general basis transformation possible just rotates
$ H_2 $ by means of a U(1) phase.
This is equivalent to rotating $ R $ and $ I $ by means of a SO(2) matrix.
Obviously,
under such a basis transformation,
the coefficient $ a $ of a term of the form 1 is invariant,
while the two coefficients $ b_1 $ and $ b_2 $
of a term of the form 2 transform as a vector of SO(2).
Similarly,
the coefficients $ c_{ij} $ of a term of the form 3 transform as a
second-rank tensor,
the coefficients $ d_{ijk} $ of a term of the form 4 constitute a
third-rank tensor,
and so on.
But the phase of $ H_2 $ is arbitrary and meaningless.
Therefore,
only quantities invariant under this basis transformation
are meaningful and can appear in the final result for any physical quantity.
For instance,
$ a $,
or $ (b_1)^2 + (b_2)^2 $,
or $ c_{11} c_{22} - (c_{12})^2 $,
are invariant.

Now consider a CP transformation.
In the simplest CP transformation,
besides the change of sign of the space coordinates,
both $ H_1 $ and $ H_2 $ transform
to their hermitian conjugates\footnote{The hermitian conjugation
arises in order to obtain the correct CP transformation
of the gauge interactions of the scalars.
In a model without gauge interactions,
the hermitian conjugation would not be necessary,
and CP conservation would be automatic.
Indeed,
CP loses its meaning if there are no gauge interactions
to distinguish the particles
from the anti-particles.}.
This sets the CP properties for the $H_1$ component fields:
$H^0$ is CP-even,
$G^0$ is CP-odd,
$G^-$ and $ G^+$ are CP conjugates.
For the $H_2$ components it dictates the transformations:
$H^- \leftrightarrow H^+$,
$ R \rightarrow R $ and $ I \rightarrow - I $.
However,
because the phase of $ H_2 $ is meaningless,
a general CP transformation must also include a phase rotation of $ H_2 $.
For instance,
$ H^- \leftrightarrow - H^+ $,
$ R \rightarrow - R $ and $ I \rightarrow I $
is also a CP transformation.
In a general CP transformation,
$ R $ and $ I $ get rotated
by an O(2) matrix with determinant $ -1 $.
A CP transformation is analogous to a basis transformation,
the only difference being that the O(2) matrix which transforms
the coefficients $ b_i $,
$ c_{ij} $,
and so on,
has determinant $ -1 $ for a CP transformation,
while it has determinant $ +1 $ for a basis transformation.

Our purpose is to find quantities which are invariant under a
basis transformation but change sign under a CP transformation.
We need for that purpose the antisymmetric tensor
$ \epsilon_{ij} $,
with $ \epsilon_{12} = 1 $.
This quantity behaves as a second-rank tensor under a basis transformation,
but changes sign under a CP transformation:
\begin{equation}
O_{ir} O_{js} \epsilon_{rs}
= (\det O) \epsilon_{ij}\, .
\label{eq:epsilon}
\end{equation}
It is easy to construct CP-odd invariants using $ \epsilon $.
For instance,
$ b_i b^{\prime}_j \epsilon_{ij} $,
$ b_i c_{ij} b_k \epsilon_{jk} $,
$ e_{ijjk} c_{kl} \epsilon_{il} $,
and so on,
are quantities which,
if they are non-zero,
imply CP violation,
while being invariant under a basis transformation.

Let us apply this general method to the mass matrix $ M $.
As we pointed out before,
that mass matrix can be divided in an invariant $ a $,
a vector $ b_i $,
and a second-rank tensor $ c_{ij} $.
The simplest CP-odd invariant which can be formed from this is
\begin{equation}
J_1 = b_i c_{ij} b_k \epsilon_{jk}
= b_i c_{i1} b_2 - b_i c_{i2} b_1
= M_{12} M_{13} (M_{22} - M_{33}) + M_{23} [(M_{13})^2 - (M_{12})^2]\, .
\label{eq:j1}
\end{equation}
This is the quantity introduced by M\'endez and Pomarol \cite{pomarol},
who however did not explain its origin.
$ J_1 $ can be expressed as a function
of the masses of the physical scalars and of the matrix $ T $
by using Eq.~\ref{eq:diagonalization}.
One obtains
\begin{eqnarray}
J_1 & = & m_i^4 m_j^2 T_{i1} T_{j1} (T_{i2} T_{j3} - T_{i3} T_{j2})
\label{eq:j1first}\\
    & = & (m_1^2 - m_2^2) (m_1^2 - m_3^2) (m_2^2 - m_3^2)
          T_{11} T_{21} T_{31}\, .
\label{eq:j1second}
\end{eqnarray}
Notice that,
$ T $ being a matrix of SO(3),
$ T_{i2} T_{j3} - T_{i3} T_{j2} = T_{k1} $ with $ (ijk) $ cyclic.

The similarity of Eq.~\ref{eq:j1second} with Eq.~\ref{eq:J} is evident.
It is easy to show,
using the identity
\begin{equation}
\delta_{ik} \delta_{jl}
= \epsilon_{ij} \epsilon_{kl} + \delta_{il} \delta_{jk}\, ,
\label{eq:identity}
\end{equation}
($ \delta_{ij} $ is the Kronecker symbol)
that any other CP-odd invariant constructed from the mass matrix alone
is proportional to $ J_1 $.
Thus,
$ J_1 $ occupies in the two-doublet model a place similar to the one of
the quantity $ J $ of Eq.~\ref{eq:J}
in the three-generation standard model.
If the masses of any two of the three physical scalars are equal,
or if any of the matrix elements of the first column of $ T $ vanishes,
there is no CP violation in the mixing of the neutral scalars.
If any two of the three non-diagonal entries of the mass matrix $ M_{12} $,
$ M_{13} $,
and $ M_{23} $ vanishes,
there is no CP violation in the mixing.

Having found that $ J_1 \neq 0 $ implies CP violation,
the question arises whether $ J_1 = 0 $ implies CP conservation.
In order to answer this,
it is convenient to express $ J_1 $ as a function of the couplings
in the potential in Eq.~\ref{eq:potential}.
{}From Eq.~\ref{eq:j1} and Eqs.~\ref{eq:M11} to \ref{eq:M13} one finds
\begin{equation}
J_1 = - 8 v^6 {\rm Im} (\lambda_5^{\ast} \lambda_6^2)\, .
\label{eq:j1third}
\end{equation}
The quantity in the right-hand side of this equation
is invariant under a rephasing of $ H_2 $ and,
being an imaginary part,
it certainly implies CP violation.
But one recognizes that there are
other quantities with the same properties:
the imaginary parts of $ (\lambda_5^{\ast} \lambda_7^2) $
and of $ (\lambda_6^{\ast} \lambda_7) $.
Therefore,
we guess that,
even if $ J_1 = 0 $ implies CP conservation in the mixing
of the neutral scalars,
there will be other CP-violating quantities,
appearing in the cubic and quartic scalar interactions,
which are independent of $ J_1 $.
This point has not been emphasized before.

Let us give a simple example.
Consider the cubic interactions of the neutral scalars with the charged scalar
given in Eq.~\ref{eq:chargedinteraction}.
As pointed out after that equation,
those interactions consist of an invariant and a vector under a rotation
of $ R $ and $ I $.
We may join the vector with the second-rank tensor
from the mass matrix to form a CP-odd invariant:
\begin{equation}
J_2 = b^{\prime}_i c_{ij} b^{\prime}_k \epsilon_{jk}
= b^{\prime}_1 b^{\prime}_2 (M_{22} - M_{33})
+ M_{23} [(b^{\prime}_2)^2 - (b^{\prime}_1)^2]
= -4 v^4 {\rm Im} (\lambda_5^{\ast} \lambda_7^2)\, .
\label{eq:j2}
\end{equation}
Similarly,
we can form another invariant from the vector $ b^{\prime} $
and the vector appearing in the mass matrix:
\begin{equation}
J_3 = b_i b^{\prime}_j \epsilon_{ij}
= M_{12} b^{\prime}_2 - M_{13} b^{\prime}_1
= 2 \sqrt{2} v^3 {\rm Im} (\lambda_6 \lambda_7^{\ast})\, .
\label{eq:j3}
\end{equation}
If we write the cubic interactions of Eq.~\ref{eq:chargedinteraction}
in terms of the physical scalars $ X_i $ in Eq.~\ref{eq:physicalscalars}
as $ H^- H^+ \sum_{i=1}^3 q_i X_i $,
we get
\begin{eqnarray}
J_2 & = & m_k^2 q_k q_i  (T_{k2} T_{i3} - T_{k3} T_{i2})
          - m_k^2 T_{k1} q_i q_j T_{i1}  (T_{k2} T_{j3} - T_{k3} T_{j2})\, ,
\label{eq:j2expression}\\
J_3 & = & m_k^2 T_{k1} q_i (T_{k2} T_{i3} - T_{k3} T_{i2})\, .
\label{eq:j3expression}
\end{eqnarray}
Thus,
$ J_2 \neq 0 $ or $ J_3 \neq 0 $ imply CP violation.

In the appendix we present a specific model
in which there is no CP violation
in the scalar mixing,
but there is CP violation in the cubic and quartic scalar interactions.

We are now in a position to discuss the cases in which $S$ is a CP-odd
monomial. We are looking for CP-odd invariants,
and therefore,
by carefully keeping track of the origin of the tensors we construct,
we will know their CP properties and can use them to define invariants
which are indeed CP odd. As an example we consider the cubic terms,
\begin{equation}
2 v G^0 H^0 \left[ ({\rm Im} \lambda_6) R +
({\rm Re} \lambda_6) I \right]\ ,
\end{equation}
This provides us with a CP-odd vector,
${\bar b}_i = 2 v ({\rm Im} \lambda_6, {\rm Re} \lambda_6)^{T} \equiv -
\epsilon_{ij} b_j / v$,
which we may combine with CP-even vectors or tensors such as
$b_i$ and $c_{ij}$ to yield CP-odd invariants.
Sometimes these invariants may be identically zero.
For instance,
\begin{eqnarray}
- v b_i {\bar b}_i
& \equiv &
\epsilon_{ij} b_i b_j = 0\ ,
\\
- v b_i c_{ij} {\bar b}_{j}
& \equiv &
b_i c_{ij} \epsilon_{jk} b_k = J_1\ .
\end{eqnarray}

\section{More than two doublets}

We now extend the method of the previous section to the case
in which there are more than two scalar doublets in the theory.
Once again,
we must work in the basis in which only one of the doublets,
$ H_1 $,
has a VEV $ v $,
and that VEV is real.
Besides $ H_1 $ there are now $ n $ more doublets in the theory,
which may be written
\begin{equation}
H_{\alpha} =
\left( \begin{array}{c}
H^+_{\alpha} \\ (R_{\alpha} + i I_{\alpha})/\sqrt{2}
\end{array} \right)\, ,
\label{eq:doublet}
\end{equation}
with the index $ \alpha $ going from $ 1 $ to $ n $.
We are free to choose a different basis for these $ n $ scalar doublets.
That freedom is expressed by the invariance
of the physics under the transformation
\begin{equation}
H_{\alpha} \rightarrow K_{\alpha \beta} H_{\beta}\, ,
\label{eq:changeofbasis}
\end{equation}
$ K $ being an arbitrary $ n \times n $ unitary matrix.
Let us write $ K $ as the sum of two $ n \times n $ real matrices,
$ K_R $ and $ K_I $,
in the following way:
$ K = K_R + i K_I $.
Let us moreover group all $ n $ fields $ R_{\alpha} $
and all $ n $ fields $ I_{\alpha} $
in a $(2n)$-dimensional vector
$ u \equiv (R_1 , R_2 , ... , R_n, I_1 , I_2 , ... , I_n)^T $.
Then,
the effect of the change of basis in Eq.~\ref{eq:changeofbasis}
on the fields $ R_{\alpha} $ and $ I_{\alpha} $
(we will consider later the charged fields)
can be written $ u \rightarrow K^{(2n)} u $,
where the $ (2n) \times (2n) $ matrix
\begin{equation}
K^{(2n)} \equiv
\left( \begin{array}{cc} K_R & - K_I \\ K_I & K_R \end{array} \right)
\label{eq:matrixk2n}
\end{equation}
is orthogonal,
as a consequence of $ K $ being unitary.
Indeed,
Eq.~\ref{eq:matrixk2n} constitutes an embedding of the group U(n)
in the group O(2n).

Similarly,
a general CP transformation is written
\begin{eqnarray}
H^+_{\alpha} \rightarrow U_{\alpha \beta} H^-_{\beta}\, ,
\label{eq:CPcharged}\\
R_{\alpha} + i I_{\alpha}
\rightarrow U_{\alpha \beta} (R_{\beta} - i I_{\beta})\, ,
\label{eq:CPneutral}
\end{eqnarray}
with $ U $ a $ n \times n $ unitary matrix.
Separating $ U $ too in its real and its imaginary parts,
$ U = U_R + i U_I $,
Eq.~\ref{eq:CPneutral} can be written
$ u \rightarrow U^{(2n)} u $,
where
\begin{equation}
U^{(2n)} \equiv
\left( \begin{array}{cc} U_R & U_I \\ U_I & - U_R \end{array} \right)
\label{eq:matrixu2n}
\end{equation}
is a $ (2n) \times (2n) $ orthogonal matrix.
Notice however that $ U^{(2n)} $ is of a different form from $ K^{(2n)} $.

Let us now generalize the tensor $ \epsilon_{ij} $ of the previous section
to a tensor $ \varepsilon $.
This is a $ (2n) \times (2n) $ matrix which is written in block form as
\begin{equation}
\varepsilon =
\left( \begin{array}{cc} 0 & 1_n \\ - 1_n & 0 \end{array} \right)\, ,
\label{eq:varepsilon}
\end{equation}
$ 1_n $ being the $ n \times n $ unit matrix.
This matrix has the properties
\begin{eqnarray}
K^{(2n)} \varepsilon \left[ K^{(2n)} \right]^T & = & \varepsilon\, ,
\\
U^{(2n)} \varepsilon \left[ U^{(2n)} \right]^T & = & - \varepsilon\, .
\end{eqnarray}

The generalization of the results of the previous section is now evident.
The coefficients of
terms linear in the fields $ R_{\alpha} $ and $ I_{\alpha} $
are grouped in vectors,
the coefficients of
quadratic terms are grouped in second-rank tensors,
and so on.
Those quantities behave as vectors,
second-rank tensors,
and so on,
under transformations of the sub-group U(n) of O(2n)
realized in Eq.~\ref{eq:matrixk2n}.
When one wants to form a CP-odd quantity
which is invariant under a basis transformation,
one just picks some of these tensors and contracts all of their indices,
leaving only two free indices $ i $ and $ j $,
which then are contracted with $ \varepsilon_{ij} $.

As an example,
consider,
in the three-Higgs-doublet model,
the mass terms of the neutral scalars,
written in the form
\begin{equation}
\frac{1}{2} \left( \begin{array}{ccccc} H^0 & R_1 & R_2 & I_1 & I_2
\end{array} \right)
M \left( \begin{array}{c} H^0 \\ R_1 \\ R_2 \\ I_1 \\ I_2
\end{array} \right)\, .
\label{eq:mass2}
\end{equation}
$ M $ is now a $ 5 \times 5 $ symmetric matrix.
Besides the invariant $ M_{11} $,
this matrix includes a vector $ b_i = M_{1,i+1} $
and a second-rank symmetric tensor
$ c_{ij} = M_{i+1,j+1} $.
The tensor $ \varepsilon $ can be written as a $ 4 \times 4 $
antisymmetric matrix,
with $ \varepsilon_{13} = \varepsilon_{24} = - \varepsilon_{31}
= - \varepsilon_{42} = 1 $,
and all other matrix elements 0.
The invariant $ J_1 $ of the previous section
can be extended to the three-Higgs-doublet model:
\begin{equation}
J_1 = b_i c_{ij} b_k \varepsilon_{jk}\, .
\label{eq:j1new}
\end{equation}
We now are able,
however,
to define other CP-odd quantities,
for instance,
\begin{equation}
J_1^{\prime} = b_i c_{ij} c_{jl} b_k \varepsilon_{lk}\, .
\label{eq:j1prime}
\end{equation}
The analogous quantity for the two-doublet model
(with $ \varepsilon $ substituted by $ \epsilon $)
can be shown,
with the help of Eq.~\ref{eq:identity},
to be equal to $ c_{ii} J_1 $,
in which $ c_{ii} $ is itself a CP-even invariant.
For more than two Higgs doublets,
however,
$ J_1^{\prime} $ is a CP-odd invariant independent of
$ J_1 $.
In general,
it is possible to construct still other independent CP-odd invariants,
of ever increasing order in the elements of $ M $.

It is interesting to write the invariant $ J_1 $
as a function of the masses of the five physical neutral scalars,
and of the mixing matrix $ T $,
which is now $ 5 \times 5 $.
We obtain an obvious generalization of Eq.~\ref{eq:j1first}
\begin{equation}
J_1 = m_i^4 m_j^2 T_{i1} T_{j1}
(\varepsilon_{ab} T_{i,a+1} T_{j,b+1})\, .
\label{eq:j1newfirst}
\end{equation}
It is important to note that it is the combination of $ T $ matrix elements
in the right-hand-side of this equation
that appears in the vertex of the $ Z^0 $
with the two physical particles $ X_i $ and $ X_j $.
Similarly, for $J_1^{\prime}$ we find,
\begin{equation}
J_1^{\prime} = m_i^6 m_j^2 T_{i1} T_{j1}
(\varepsilon_{ab} T_{i,a+1} T_{j,b+1}) - (m_k^2 T_{k1}^2) J_1\, .
\label{eq:j1newprimefirst}
\end{equation}
Eqs.\ \ref{eq:j1newfirst} and \ref{eq:j1newprimefirst}
actually hold for any number of Higgs doublets.

\section{Extension to the charged scalars}

The extension to the charged scalars of the work of the two previous sections
is fairly obvious.
As is seen in Eq.~\ref{eq:changeofbasis},
the charged scalars $ H_{\alpha}^+ $ transform
under basis transformations as the $ R_{\alpha} + i I_{\alpha} $.
Together with Eq.~\ref{eq:CPcharged},
this indicates that the procedure to follow must be
to separate $ H^+_{\alpha} $ into its real and imaginary parts,
and to treat the real part as $ R_{\alpha} $ and the imaginary part
as $ I_{\alpha} $.

For instance,
the mass matrix of the charged scalars,
$ N $,
is a $ n \times n $ hermitian matrix,
appearing in the mass terms
\begin{equation}
\left( \begin{array}{cccc} H_1^- & H_2^- & ... & H_n^- \end{array} \right)
N
\left( \begin{array}{c} H_1^+ \\ H_2^+ \\ ... \\ H_n^+ \end{array} \right)\, .
\label{eq:chargedmass}
\end{equation}
After separating each field $ H_{\alpha}^+ $
into its real and its imaginary part,
and putting all the real parts, followed by all the imaginary parts,
in a $ (2n) $ vector,
these mass terms may be written using instead a $ (2n) \times (2n) $
symmetric matrix which is,
in block form,
\begin{equation}
\left( \begin{array}{cc} {\rm Re} N & - {\rm Im} N \\
{\rm Im} N & {\rm Re} N \end{array} \right)\, .
\label{eq:chargedmass2}
\end{equation}
This matrix effectively constitutes a second-rank tensor
which we may call $ h_{ij} $.
{}From this tensor,
together with the tensor and the vector occurring in the mass matrix
of the neutral scalars,
it is possible to construct further CP-odd quantities,
like
$ h_{ij} c_{jk} b_k b_l \varepsilon_{il} $,
$ c_{ij} h_{jk} b_k b_l \varepsilon_{il} $,
or
$ c_{ij} c_{jk} h_{kl} c_{lm} \varepsilon_{im} $,
and so on.

We emphasize that the treatment of the charged sector
delineated in this section
is necessary even in the two-Higgs-doublet model.
Since the mass term for only one charged boson is of the form $ H^- H^+ $,
and this is invariant under a rephasing of $ H^+ $,
the mass term of the charged scalar does not have any influence
on CP violation.
However, to deal with terms involving $ G^- H^+ H^0 $ and
$ G^- H^+ \xi_i $, we must use our methods for the charged sector.
The procedure consists in breaking
$G^\pm$ into their real and imaginary component
fields $G_R$ and $G_I$
($G^\pm = G_R \pm i G_I$),
and noting that they have CP even and CP odd properties,
respectively.

For example, in the cubic potential we have terms such as
\begin{equation}
\sqrt{2} v G^- H^+ \left[ \frac{\lambda_4}{2} (R - i I) +
\lambda_5 (R+ i I) + \lambda_6 H^0 \right] + h.c.\ .
\end{equation}
Using $ \xi_1^c = {\rm Re} H^+ $ and
$ \xi_2^c = {\rm Im} H^+ $, we can rewrite
this as,
\begin{equation}
G_R H^0 g_i \xi_i^c + G_R \xi_i^c k_{ij} \xi_j + \cdots\ ,
\end{equation}
where we have suppressed the analogous CP odd terms in $G_I$ and used,
\begin{eqnarray}
g_i & = & \frac{\sqrt{2}}{v} b_i\ ,
\\
k_{ij} & = & \frac{2 \sqrt{2}}{v} [ c_{ij} -
\frac{1}{2}
(\mu_2 + \lambda_3 v^2) \delta_{ij}]\ .
\end{eqnarray}
Therefore, we can recover $J_1$ by considering transformations of the
charged fields.

\section{Discussion and conclusions}

In this paper we addressed the problem of the systematic construction
of CP-odd quantities for a SU(2)$\otimes$U(1) model
with many scalar doublets.
We pointed out that basis trasformations leaving
the VEV $ v $ invariant have no physical meaning.
Therefore,
our strategy has been to define CP-odd quantities
which are invariant under those basis transformations.

Our method does not discriminate between the two possible origins
of CP violation,
explicit or spontaneous.

Our method does not rely on the assumption that
the scalar potential is quartic.
It can therefore be directly applied in effective scalar theories,
with non-renormalizable interactions.

The way is now open to search for experimental signatures
of CP violation in the scalar sector \cite{norway,pomarol},
and to determine which basic CP-violating quantities
we might be able to measure.

\section*{Appendix}

In this appendix we present a model with a symmetry
such that there is no CP violation
in the scalar mass terms,
all the CP violation being in the cubic and quartic interactions.

Take three scalar doublets $ H_1 $,
$ H_2 $ and $ H_3 $,
with a symmetry $ \Delta(27) $ \cite{gerard,kaplan}.
This is the discrete subgroup of SU(3) generated by the two transformations
\begin{equation}
H_1 \rightarrow H_2 \rightarrow H_3 \rightarrow H_1
\end{equation}
and
\begin{equation}
H_1 \rightarrow H_1 ,\ H_2 \rightarrow \omega H_2 ,\
H_3 \rightarrow \omega^2 H_3\, ,
\end{equation}
where $ \omega = \exp (2 \pi i / 3) $.
The most general quartic Higgs potential invariant
under this symmetry is
\begin{eqnarray}
V & = &
\mu (H_1^{\dagger} H_1 + H_2^{\dagger} H_2 + H_3^{\dagger} H_3 )
\nonumber\\
  &   &
+ a \left[ (H_1^{\dagger} H_1)^2 + (H_2^{\dagger} H_2)^2
+ (H_3^{\dagger} H_3)^2 \right]
\nonumber\\
  &   &
+ b \left[ (H_1^{\dagger} H_1) (H_2^{\dagger} H_2)
+ (H_1^{\dagger} H_1) (H_3^{\dagger} H_3)
+ (H_2^{\dagger} H_2) (H_3^{\dagger} H_3) \right]
\nonumber\\
  &   &
+ c \left[ (H_1^{\dagger} H_2) (H_2^{\dagger} H_1)
+ (H_1^{\dagger} H_3) (H_3^{\dagger} H_1)
+ (H_2^{\dagger} H_3) (H_3^{\dagger} H_2) \right]
\nonumber\\
  &   &
+ \left\{ d \left[
(H_1^{\dagger} H_2) (H_1^{\dagger} H_3)
+ (H_2^{\dagger} H_1) (H_2^{\dagger} H_3)
+ (H_3^{\dagger} H_1) (H_3^{\dagger} H_2) \right] + h.c. \right\}\, .
\end{eqnarray}

One possible vacuum of this potential has $ H_1 $ with VEV $ v $
while both $ H_2 $ and $ H_3 $ have vanishing VEV.
Then,
the only mass terms with a complex phase originate in
$ d (H_1^{\dagger} H_2) (H_1^{\dagger} H_3) $,
because $ d $ is complex.
But it is obvious that we may redefine the phases of $ H_2 $ and $ H_3 $
in order to absorb the complexity of $ d $ in that mass term.
On the other hand,
that rephasing of $ H_2 $ and $ H_3 $ does not lead to the elimination
of the complex phases in the terms
$ d (H_2^{\dagger} H_1) (H_2^{\dagger} H_3) $ and
$ d (H_3^{\dagger} H_1) (H_3^{\dagger} H_2) $,
which yield cubic and quartic interactions when $ H_1 $ gets its VEV.
Therefore,
in this natural model there is no CP violation
in the mixing of the scalars,
but there is CP violation in their cubic and quartic interactions.

Notice that this is what happens too in the model with $ \Delta(27) $
symmetry of Branco, Grimus, and G\'erard (BGG) \cite{gerard}.
There,
the coupling $ d $ is real and CP is spontaneously broken,
when $ H_1 $,
$ H_2 $ and $ H_3 $ acquire VEVs with relative phases.
However,
in that model,
when one goes to the basis in which only $ H_1 $ has
a non-vanishing VEV,
one obtains a potential which still has $ \Delta(27) $ symmetry,
but now with a complex $ d $.
Therefore,
the model of BGG has CP violation only in the cubic and quartic
scalar interactions,
but not in the mixing of the scalars.

\vspace{2mm}

We thank L.\ Wolfenstein for discussions.
Both he and Ling-Fong Li read the manuscript.
This work was supported by the United States Department of Energy,
under the contract DE-FG02-91ER-40682.
J.\ P.\ S.\ was partially supported by the Portuguese JNICT, under
CI\^{E}NCIA grant \# BD/374/90-RM, and he is indebted to the Santa
Barbara Institute for Theoretical Physics,
where this work has been started.

\vspace{5mm}

%
%

\begin{thebibliography}{99}
%
\bibitem{jarlskog}
C.\ Jarlskog,
Phys.\ Rev.\ Lett.\ {\bf 55}, 1039 (1985).
%
\bibitem{wu}
I.\ Dunietz, O.\ W.\ Greenberg, and D.-D.\ Wu,
Phys.\ Rev.\ Lett.\ {\bf 55}, 2935 (1985).
%
\bibitem{branco}
J.\ Bernab\'eu, G.\ C.\ Branco, and M.\ Gronau,
Phys.\ Lett.\ {\bf 169B}, 243 (1986).
%
\bibitem{gronau}
M.\ Gronau, A.\ Kfir, and R.\ Loewy,
Phys.\ Rev.\ Lett.\ {\bf 56}, 1538 (1986).
%
\bibitem{lavoura}
G.\ C.\ Branco and L.\ Lavoura,
Nucl.\ Phys.\ {\bf B278}, 738 (1986).
%
\bibitem{rebelo}
G.\ C.\ Branco and M.\ N.\ Rebelo,
Phys.\ Lett.\ B {\bf 173}, 313 (1986).
%
\bibitem{nesbitt}
G.\ C.\ Branco, L.\ Lavoura, and M.\ N.\ Rebelo,
Phys.\ Lett.\ B {\bf 180}, 264 (1986).
%
\bibitem{kostelecky}
G.\ C.\ Branco and V.\ A.\ Kostelecky,
Phys.\ Rev.\ D {\bf 39}, 2075 (1989).
%
\bibitem{grimus}
G.\ Ecker, W.\ Grimus, and W.\ Konetschny,
Nucl.\ Phys.\ {\bf B191}, 465 (1981);
G.\ Ecker, W.\ Grimus, and H.\ Neufeld,
Nucl.\ Phys.\ {\bf B247}, 70 (1984);
G.\ Ecker, W.\ Grimus, and H.\ Neufeld,
J.\ Phys.\ {\bf A20}, L807 (1987);
H.\ Neufeld, W.\ Grimus, and G.\ Ecker,
Int.\ J.\ Mod.\ Phys.\ A {\bf 3}, 603 (1988);
%
\bibitem{pomarol}
A.\ M\'endez and A.\ Pomarol,
Phys.\ Lett.\ B {\bf 272}, 313 (1991).
%
\bibitem{gerard}
G.\ C.\ Branco, J.-M.\ G\'erard, and W.\ Grimus,
Phys.\ Lett.\ B {\bf 136}, 383 (1984).
%
\bibitem{kaplan}
A recent paper with a good introduction to the $ \Delta $
(dihedral) symmetries is
D.\ B.\ Kaplan and M.\ Schmaltz,
Phys.\ Rev.\ D {\bf 49}, 3741 (1994).
%
\bibitem{norway}
A.\ Skjold and P.\ Osland,
University of Bergen Report No.\ ISSN 0803-2696 (1994),
unpublished.
%
%
\end{thebibliography}
\end{document}